\documentclass[runningheads]{llncs}
\usepackage{graphicx}
\usepackage{float}
\usepackage[caption=false]{subfig}
\usepackage{url}
\usepackage{todonotes}
\usepackage{xcolor}
\usepackage{float}
\usepackage{subfig}
\usepackage{algorithm}
\usepackage{algpseudocode}

\begin{document}

\mainmatter              
\title{Leader Election in the Internet of Things: Challenges and Opportunities}
\titlerunning{}  

\author{Mohsin Ur Rahman 
}

%
\tocauthor{Mohsin Ur Rahman}
\institute{Department of Computer Science, University of Pisa\\
\email{\{mohsinur.rahman\}@di.unipi.it}}

\maketitle
\thispagestyle{empty}
\pagestyle{empty}
%
%

%
%
%
\begin{abstract}
The different functions of a leader node in Ad hoc networks, particularly in Wireless Sensor Networks (WSNs) and Internet of Things (IoT) include generation of keys for encryption/decryption, finding a node with minimum energy or node located in an extreme side of the network.  One essential application of IoT is to monitor any dangerous, sensitive or non-accessible site. Such type of application requires the election of a leader located on the extreme left of the network. Indeed, one can then use any leader election (LE) algorithm to find this node, and then start the process of finding its boundary nodes. However, the chosen algorithm must be fault-tolerant and robust since the detection of a node failure under such circumstances becomes impossible. This paper surveys the most significant work performed in this direction, highlights the challenges and proposes possible extensions.

\end{abstract}
\section{Introduction}
The emergence of new technologies such as radio-frequency identification, wireless sensor networks, and short-range wireless communication has enabled the Internet to penetrate in embedded computing. The Internet of Things (IoT) \cite{IoT}, which will allow the interconnectivity of everyday objects, is now becoming a reality because daily physical objects are increasingly becoming equipped with different types of (uniquely) addressable sensors, thereby allowing the connectivity with them through the Internet. A successful effort was made to port the IP stack on these embedded devices, together with the emergence of IPv6 (which offers enormous addressing space), eased merging of the digital and physical world, hence enabled the IoT to advance faster.

In the context of IoT, the useful applications of Wireless
Sensor Networks (WSNs) include situations where the sensors
need to be deployed in hazardous and inaccessible locations.
Furthermore, they are also proving useful to detect and prevent
intrusions around sensitive sites. A WSN usually
consists of many sensor nodes that use short wireless communication to exchange information with each other. Nowadays, WSNs and IoT both support the execution of many other popular distributed applications as well, e.g., data aggregation, group communication, encryption/decryption operations, exchange of data, discovery of boundary nodes and agreement problem. The execution of such critical applications, particularly, in WSN-like fault prone environment, the elected leader plays a critical role in order to coordinate such tasks in a distributed fashion.

A WSN is easily installable and, thus, offers a popular and cheap computing infrastructure; however, it is also a challenging and constrained environment. For instance, a WSN node usually has low storage, memory and computing capabilities. Furthermore, each node can communicate with the other node via message passing over short range wireless links. Nodes located outside of each other communicate range can only exchange information via message relay. It is interesting to observe that arbitrary topological changes may occur due to non-deterministic mobility patterns. In addition, more frequent topological changes occur due to the dynamism in wireless links. Thus, this situation seriously affects the performance of the network in terms of message delay. Hence, the presence of (movable) sensor nodes requires the development of new distributed algorithms because the algorithms already developed for static scenarios are not directly implementable in such environments.

Processes are the fundamental elements of a distributed systems relying on either asynchronous or synchronous communication to exchange information with each other. However, these processes require the selection of a common leader so that it can perform some essential tasks on their behalf, such as control and coordination. Therefore, selecting a unique leader to perform such tasks is a challenging problem, termed as the {\em leader election (LE)} problem. The algorithms presented in this survey are primarily concerned with one of the critical applications of IoT to find the boundary nodes in order to monitor sensitive, inaccessible and dangerous sites.

The LE problem can not be solved in anonymous systems (i.e., systems which do not permit the identifiers of nodes). In other words, the coordination or control tasks require the differentiation among the different processes in the system. Indeed, this can only be accomplished when each
node possesses a {\em unique identifier}. Therefore, existing LE algorithms in IoT assume that each process $p_i$ has a unique identifier, e.g., id. Moreover, these algorithms also assume that the identifiers are comparable with each other.

An algorithm can be classified into three categories depending on the network synchronization level that it demands \cite{Kuhn} \cite{Tseng}.  {\em Synchronous} algorithms rely on the assumption that the
message delay is bounded, and processes have access to a
global clock or they use synchronous rounds to operate. {\em Partially asynchronous} rely on the assumption that the message delay is bounded. On the other hand, {\em fully asynchronous algorithms} rely on the assumption
that either nodes in the network may halt for an unspecified
amount of time or they assume an unbounded message delay.
Please note that the problem of leader election cannot be solved in fully asynchronous networks. Therefore, the ability of these algorithms to determine the weakest assumption in terms of changes in the network, and, particularly, synchronization is still an open issue \cite{Larrea}.

\section{Leader Election Algorithms}
In the literature, two main families of LE techniques can be identified. The first one uses the {\em ring topologies} for leader election while the second one is based on arbitrary topologies (i.e., ad hoc networks). It is worthy to note that leader election in IoT is based on tree topologies because they are proving useful to solve other problems as well. In the following section, we discuss LE algorithms for both ring as well as ad hoc topologies highlighting their advantages and limitations. In this section, we provide an overview of the state-of-the-art as well as existing techniques that can be used to further enhance the performance of the current LE algorithms in IoT.

Several clustering algorithms are proposed for WSNs to
achieve different features such as fault-tolerant and energy
consumption. Thus, we believe that the work done in the
aforementioned domain can greatly help in further improving the performance
of the current LE algorithms in IoT.

\subsection{Leader Election for Ring Topologies}
An interesting layered architecture to find the leader node in Mobile Ad Hoc Networks is recently proposed in \cite{sharma}. At the lowest layer, clustering algorithms are applied to divide the network into balanced clusters. To this end, they have exploited the merging clustering algorithm (MCA) \cite{MCA} in order to construct a heap of cluster nodes. The second layer then exploits the ring formation algorithm (RF) which forms the virtual ring of cluster heads (CHs). This layer uses a heap data structure to represent members in the ring. Finally, an LE algorithm is applied to elect one of the CHs as the leader in the network. However, a participant node may not become a part of the computation because topological changes may occur due to forced termination and power off. To deal with this problematic, they have introduced a fault-tolerant module to recover from node-crashes.

Another simple algorithm to find the group leader for the ring topology is presented in \cite{Hirschberg}, where the authors assume the case of a {\em bidirectional rings}, allowing a process to transmit and receive election messages from any neighbor located on any side. Another algorithm for {\em unidirectional rings} is proposed in \cite{Dolev}, which uses FIFO mode for channels to send and receive messages. The solution proposed in \cite{Hirschberg} uses different rounds in order to allow processes to compete for leader election. In the first round, all processes compete to be selected as leaders in the networks, but only half of them becomes eligible for the next round. The next rounds also uses a similar strategy until a {\em global} unique leader is elected in the system.

\subsection{Leader Election for Ad Hoc Topologies}
A good and efficient LE algorithm possesses
some special capabilities such as offering low communication
overhead, low message and time complexity, and the longevity
of the elected leader. The
Bully Algorithm \cite{Effat} is the classical algorithm to deal with this
issue in synchronous systems. Indeed, the algorithm requires
a large number of election messages to elect a unique leader
in the network. It is worthy to note that this algorithm is not suitable for
real-time IoT applications due to its high number of message
passing in the system.
The basic objective of the Bully algorithm is that all
processes agree to select a highest process ID or process with
the highest priority as the common coordinator (i.e., leader). However, the time and message complexities of this algorithm are both $O(n^2)$.

\begin{figure}[H]
  \centering
  \includegraphics[scale=0.65]{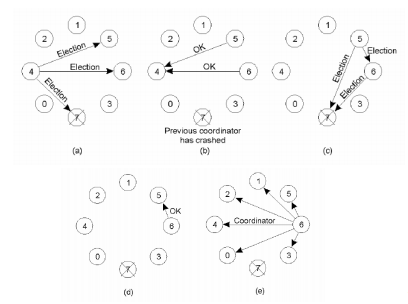}
  \caption{Illustration of the Bully Algorithm}
  \label{new}
\end{figure}

The LE process using the Bully Algorithm is
explained in Figure 1. As it can be observed, process 4 initiates the election message. However,
process 4 would receive an OK message to stop the election
because of the existence of higher ID processes in the system.
Therefore, process 5 and 6 respond and inform 4 to stop. After
that, two other processes (i.e., 5 and 6) initiate individual
elections. However, process 6 inform 5 to stop by sending
an OK message (see Fig. 2 (d). Thus, process 6 wins the
election because it did not receive response from other higher
processes. Thus, it, finally,  informs all other processes that it is the new
coordinator.

A {\em performance-based LE algorithm} is presented in \cite{Kim}. They have first divided the network, and claim that an optimal leader is the nearest node from all nodes and has better performance compared to the other nodes in the system.  Thus, the pre-election algorithm allows the election of a leader based on the performance and operation rate of nodes and links. Therefore, each node is capable to detect the failure of the current leader, and, consequently, replaces it with the provisional leader. Indeed, this criterion is used to improve the predictability of pre-election algorithm (i.e., it becomes easier to suddenly choose a provisional leader based on its operation rate). Main advantage of the proposed algorithm include adaptability to any kind of network topology. However, this algorithm requires the existence of reliable entities in the system. This algorithm requires $O(k(k - D)$ time complexity to elect a leader while its message complexity is $O(n + k)$, where D indicates the diameter of the distributed network, and k denotes nodes that are operating $(k\leq 5)$.

 The authors in \cite{Effat} have proposed an improved version of the Bully algorithm in order to further reduce the number of exchanged messages. A drawback of the existing Bully algorithm is that the message carrying the highest node ID may get lost before reaching N.  Therefore, this modified algorithm  considers fault-tolerant capabilities in order to prevent the loss of LE message. The basic steps of this algorithm are similar to \cite{Soundarabai}  except for the provision of this new functionality. The message complexity using this algorithm is greatly reduced to $log n + (n-1)$. Another modification of the Bully's algorithm is proposed in \cite{Park}, where the LE criteria is adapted from \cite{Kim}.

  An efficient distributed algorithm for LE in bounded-degree networks is proposed in \cite {Chow}. This algorithm achieves improved performance by limiting the number of nodes which can be used to detect the failure of nodes. Thus, even in worst-case scenarios, the algorithm achieves optimal time and message complexities. According to their network model, the time and message complexities of the proposed algorithm are optimal (i.e., $O(n)$).

  The authors in \cite{Rafailescu} emphasize the importance of the leader in distributed systems. However, the existence of the leader node is not always guaranteed due to technical problems or crashes. They claim that such scenarios need a clear procedure in order to select a new one. Furthermore, satisfying some random conditions to elect a new leader are hard because all the nodes must be given equal chances. Thus, they have exploited a {\em random roulette wheel} selection technique to select a new leader from the remaining nodes in a fault tolerant manner. (1)	The algorithm requires a node to satisfy the launch condition before it becomes eligible to execute the roulette wheel technique. The launch condition refers to a situation in which a node generates a random number in (0,1) interval, but greater than a selected threshold (i.e., $>0.80$). Thus, a node matching this condition then becomes an eligible candidate, and it then uses this number to send a special message to all other nodes (2) The node receiving this message is required to vote for the sender (3) The coordinator will be the node which has received all the positive votes (4) After deciding the leader, the coordinator inform it by sending a message, and, finally, the newly selected leader broadcasts a special/heartbeat message. This algorithm needs 2n -1 messages (excluding the final heartbeat messages).

\section{Leader Election in IoT}

The problem of finding a unique leader in IoT mainly relies on solving the problem of finding a minimum spanning tree (MST). However, finding an {\em optimal distributed algorithm} to construct MST is the main challenge to efficiently solve the problem of LE in IoT. In other words, efficient solutions demand an efficient technique to construct the MST. It is interesting to observe that trees are suitable for LE in IoT since these are also helpful for other tasks including deadlock resolution, network synchronization, and for performing breadth-first search. In addition, MST minimizes communication overhead by allowing the information to be forwarded over a minimum weight spanning tree \cite{Awerbuch}.

Tree-based LE algorithms in IoT are primarily concerned with constructing MST in a distributed fashion. To achieve this objective, a few algorithms are proposed in the literature \cite{Bounceur1} \cite{Bounceur2} \cite{Bounceur3}. These algorithms usually assume that each user node is identified by its unique identifier (e.g., id which is an integer number). Furthermore, edges form the logical links among nodes where each edge has an associated cost or weight. The smallest connected components form sub-trees where the root process is used to uniquely represent each sub-tree. Moreover, each sub-tree gathers information about edges and vertices of its associated nodes. As a result, a big tree known as minimum spanning tree (MST) is formed by combining the small sets of logical sub-trees.

Informally, we are interested in a self-stabilizing system that allows recovery from any arbitrary global state to reach a preferred global state within finite time. Moreover, this preferred global state is always constant in
terms of program execution, i.e., the execution of a program
can not alter the stable state once it is reached. Thus, we
believe that the work done in the self-stabilizing system can
greatly help in solving the LE problem in IoT.
Furthermore, a few self-stabilizing LE algorithms
are recently presented in \cite{Datta} \cite{Altisen} \cite{Blin}. However, these algorithms are
not suitable for message-passing systems since they assume a
shared-memory model. It is interesting to note that both self-stabilizing and terminating solutions are not admittable when
considering spanning tree and the LE problem. A
terminating algorithm is the one that touches a fixpoint state
by disabling all program actions [25].

The LE problem requires the existence of a unique leader in the system, and a correct solution must guarantee this condition
[45]. It is interesting to observe that this condition
requires that each process possesses a unique identifier. Otherwise, it is impossible to select a global unique leader in the
system as it is already proven. However, guaranteeing a unique
leader at all times is impossible [42]. For example, some of the
components will lose their leader if the network is partitioned
into groups. Similarly, merging components together has an
opposite effect (i.e., two or more leaders will then serve the
merged components, which does not guarantee the uniqueness
condition discussed above). We now discuss the LE algorithms in IoT highlighting their strengths and weaknesses.

\subsection{Minimum Finding Algorithm (MFA)}

 The MFA algorithm proposed in \cite{Bounceur3} exploits the tree-based broadcast technique to elect a leader node.  This algorithm uses a few steps to find the leader node with the minimum x-coordinate in the network. First, each node in the network assigns its local value (i.e., x-coordinate) to a variable. It then broadcasts its own value and observes the local values of its neighboring nodes. If it receives a value smaller than its own, then it updates it own variable with the received value, and, consequently, broadcasts the updated value. Therefore, this process guarantees that a unique sensor node with the minimum x-coordinate will not receive a minimum value than its own. This node is, finally, selected as the leader node in the network. However, the proposed method is not proved to be secure against different types of spoofing attacks \cite{spoofing}. In addition, this algorithm requires $O(n^2)$ messages to elect a new leader in the network.


\subsection{DoTRO}
The {\em Local Minima Finding Algorithm (LMF)}  algorithm \cite{Bounceur3} uses the same principle as the MinFind algorithm to find the local minimum in the network, which indicates the node having no neighbor. However, this algorithm allows nodes to send their coordinates only once, and the decision to determine the local minimum takes place after a node has collected  x-coordinate messages for all of its neighboring nodes. Please note that each node itself decides weather it is a local minimum or not depending on the received values from its neighbors. Thus, this process can be easily falsified because a malicious node can send false information in order to deceive the other nodes in the network. This algorithm requires $O(n+1)$ messages to determine the local minima in the network.

The DoTRO algorithm \cite{Bounceur3}  first uses the LMF algorithm to find the local minima in the network.  After that, each local minima node initiates the flooding process in order to route its value over the tree.  Thus, two or more spanning tree may meet each other, and the one with the best (i.e., minimum) value continues the process while the other is suspended. Finally, the root of the final dominating tree is selected as the global minima (i.e., leader) in the network. This algorithm elects a leader in a fault-tolerant fashion, however, it also requires $O(n^2)$ messages to complete its operation.

\subsection{BROGO}
The BROGO algorithm \cite{Bounceur1} uses a flooding technique to construct the spanning tree, and to determine all possible leaf nodes in the network. Once this process is completed, each leaf broadcasts a message to a (single) reference node that contains information about the neighboring nodes. After receiving messages from all leaf nodes, the reference node then compares them to choose the leader node. Finally, the reference node informs the elected leader by sending a message confirming that it is the elected leader. However, this algorithm is vulnerable to single point of failure because it is based on a single reference node. Thus, the election process would not be successful if this node is unavailable due to possible failure. In addition, it does not elect a leader in a fault-tolerant manner. It is worthy to note that a {\em movable} reference node can further minimize energy consumption and improve network lifetime as it is already proven \cite{Kumar}.

\subsection{Revised BROGO}
To overcome the reliability issue of the BROGO algorithm, a
revised BROGO algorithm with the same message complexity as BROGO is proposed in \cite{Bounceur3}, which is based
on the WBS (Wait-Before-Starting) concept proposed in \cite{wbs}.
They also assume the minimum value for election purposes. It
is interesting to note that any node has to wait for some time
(i.e., $t = x \times w$, where x represents a node's identifier and
w a given waiting time unit) before initiating the algorithm.
Therefore, a sufficiently high value of w is required so that all
the nodes are informed about the termination of the existing
root. Thus, a second root node will start the process if the
first one fails, a third will start it if the second one fails and
so on. However, the WBS concept can be further improved for non-integer values (i.e., performance-based parameters) by multiplying the value by a certain value and then transforming the obtained values to an integer value. 

\section{Challenges and Open Issues}
In this section, we briefly survey research aimed at resolving these challenges and identify several open issues involved in providing the critical features of these algorithms (i.e., fault-tolerance, data aggregation, mobility, quality of service and security and privacy).
\subsection{Fault Tolerance}
The occurrence of faulty nodes and noisy communication media in the IoT domain make it necessary to design fault-tolerant solutions \cite{Dolev}. The concept of fault tolerance is strongly related to self-stabilization. Roughly speaking, a self-stabilizing LE algorithm is designed in such a way that it can easily cope with the failure of a leader in a system. Such algorithms usually elect a new leader when the existing one fails. However, it should be noted that stabilizing the system requires that the intermediate period between one recovery and the next faulty period should be long enough so that all the nodes are informed about the existing leader to finish \cite{Bounceur2}.

\begin{itemize}
\item LE algorithms require well-structured and coherent techniques to fault tolerance in order to ensure that these techniques will not reduce, rather than increase, the reliability of the system.
\item  When considering any LE algorithm, {\em synchronization} plays an important role in terms of overall network performance. Schemes such as slotted TDMA enables nodes to periodically schedule sleep intervals in order to minimize energy consumption. Synchronization is also necessary to ensure that an assistant node can take over the charge if a leader node fails.
\end{itemize}

\subsection{Data Aggregation}
WSN is a vital part of IoT, and the investigation of efficient data gathering techniques is essential in order to further minimize energy consumption \cite{Mondal}. This technique to eliminate {\em unnecessary information} and to find only the {\em appropriate data} for transmission is called data aggregation. Thus, aggregator nodes in IoT employ data aggregation techniques, which are mainly associated with different data aggregation functions such as SUM, COUNT, MAX, MIN, AVG, etc. to find only suitable information by summarizing and integrating data from different nodes. Furthermore, efficient data aggregation techniques demand aggregating together the correlated  data packets defined by various sensors such as humidity sensors, position sensors, temperature sensors and so on. We introduce the data aggregation architecture for the LE algorithms in IoT as shown in Figure 2.

Our architecture shows two data models-- the first one does not use aggregation while the second uses an aggregator node (i.e., node 8) to forward the collected data to the reference node. It is interesting to observe that node 4 forwards all the data packets to the reference node since it does not employ any data aggregation algorithm. On the other hand, node 8 uses an aggregation algorithm and forwards only a single packet to the reference node in order to remove duplicate data and to minimize the number of transmissions, which in turn reduces energy consumption and improves network lifetime. A systematic literature review discussing the strengths and weaknesses of the various data aggregation techniques in IoT is recently presented in \cite{Pourghebleh}.

\begin{figure}[H]
  \centering
  \includegraphics[scale=0.4]{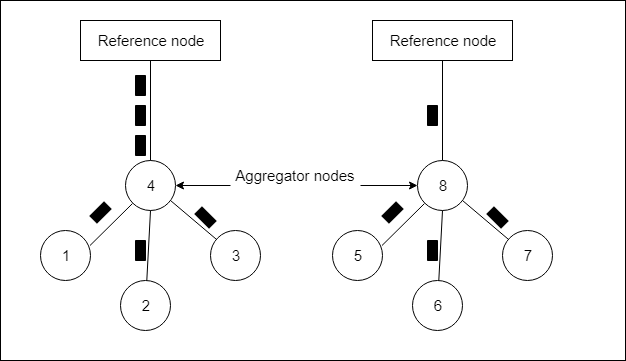}
  \caption{Data Aggregation Architecture for LE Algorithms in IoT}
  \label{fig}
\end{figure}

\subsection{Mobility}
The mobility of nodes can break the linkages
of nodes \cite{Bello}. Therefore, we believe that such algorithms
should be adaptive to arbitrary topological changes depending
on the different applications of (movable) sensors, and dynamic sensor networks \cite{dynamic} \cite{rahman2016simulation} \cite{rahman2016investigating}. In other words,
it becomes essential for other nodes to detect such changes
and to take necessary actions. In particular, these algorithms should define new
mechanisms to support {\em localization with mobility} \cite{Rawat}, to manage {\em network partitioning}, and the consequences of {\em merged partitioning} \cite{Vasudevan}.

\begin{itemize}
\item Mobility can affect the performance of LE algorithms in terms of the number of election messages in the network. For instance, the authors in \cite{sharma} shows  that link failures can occur due to the high speed of nodes in the network.  An interesting future work would be to evaluate the performance of the LE algorithms in IoT by taking nodes' movement into consideration.

\item Clustering algorithms need further investigation because the effects of (inter-cluster) and (intra-cluster) scenarios are not yet evaluated for LE algorithms in IoT.

\item Multi-leaders can be elected in order to further minimize energy consumption and to prolong network lifetime. The election of multiple leaders is particularly suitable for dynamic environments; supporting the mobility of nodes in the network \cite{Yu}.

\end{itemize}

\subsection{Quality of Service}
When designing such algorithms for IoT, the QoS requirements should also be taken into consideration. Existing LE algorithm for IoT focus on energy consumption, but pay less attention to this metric except the scheme recently proposed in \cite{Bounceur3}.

\begin{itemize}
\item Chen et al. \cite{Chen} first proposed a set of failure detectors with a model for QoS. Thus, the concept of QoS for failure detectors was initially proposed in this work, together with quantitative techniques to measure it. Roughly speaking, QoS means that the failure detector is configured in such a way that it can operate according to the network probabilistic behavior, and according the to requirements imposed by the application (i.e., packet loss tolerance).

\item {\em Scalability}  is also an important factor when considering the QoS requirements. For instance, some network architectures may handle a large number of nodes, however, such networks can seriously affect the Quality  of Service when the {\em number of nodes} are increased in the network \cite{sharma}. It is worthy to note that the effects of scalability and QoS are not extensively studied and evaluated for the LE algorithms in IoT. In addition, the performance is not evaluated by varying {\em network area }, which is also an essential factor to determine the trade-off between scalability and QoS.

 \item An interesting application is recently proposed in \cite{Reis} which meets the desired QoS requirements. It exploits counter of uptime to {\em prioritize stable processes} while relying on a single write in stable storage so that the sequence of heartbeats is not interrupted by crashes, thus, seems to be {\em paused}, which in turn reduces the chances of crash failures.

\item The authors in \cite{Ni} insist that {\em topology control} is one of key issue of sensor networks, which is essential to reduce communication problems and to prolong network lifetime. To this end, several clustering and optimization techniques  are proposed to further improve network lifetime. We believe that the work done in this direction can further improve the performance of the current LE algorithms in IoT.

 \end{itemize}

\subsection{Security and Privacy}

Security analysis of these algorithms show that they are vulnerable to a number of security and privacy issue. However, we first provide an overview of the security characteristics of IoT before discussing these vulnerabilities.

\begin{itemize}
\item	Confidentiality: Confidentiality is used to ensure that only authorized users can access the data. It is an essential feature because devices in IoT broadcast sensitive measurement information, which must be protected from unauthorized users in the network.
\item	Availability: Availability is used to ensure that the data and devices in IoT are always available for authorized users. For instance, the reference node must be always available to ensure the success of the LE process.
\item	Integrity: Integrity is used to guarantee that the data is not altered during transmission. For instance, erroneous LE messages can seriously disrupt the operations of IoT applications.
\item	Identification and Authorization: Identification is used to ensure that only authorized users are present in the network. Authentication is used to ensure the legitimacy of the delivered data in IoT. For example, these   algorithms must identify weather the LE messages are originating from authorized nodes in IoT.
 \item Privacy:  This feature is used to ensure that the receiving nodes only possess {\em limited control} on the received data, thus, not allowing them to process or access the data  \cite{Yang} \cite{Lopez}. We observed several limitations of the LE algorithms in terms of security and privacy, which are discussed in the following section.
\end{itemize}

\begin{itemize}

\item Replay Attack: An adversary can exploit malicious nodes to fraudulently delay or repeat a valid data transmission \cite{Tewari}. To overcome this issue, techniques such as {\em precise clock synchronization} and/or {\em timestamp} are required \cite{Li}.

\item Sybil Attack: A malicious node can launch Sybil attack to claim {\em multiple legitimate identities} in the network and use them for impersonation purposes in IoT \cite{Adat} \cite{rahman2017lightweight}. Thus, LE algorithm needs to consider the consequences of such attack since false data can be easily accepted by the reference node(s). Indeed, {\em signal strength-based techniques} and/or {\em secure identification mechanisms} are needed to overcome this issue in IoT \cite{Jan}.

\item	Spoofing Attack: An adversary can launch spoofing attack to gain full access to the IoT system \cite{Lu}. For instance, {\em IP spoofing}  can be used to spoof the reference node or the elected leader. After that, it can send malicious messages in the system, thus, enabling the attacker to gain complete control over the different applications of LE. Similarly, an attacker can also launch {\em RFID spoofing}  in order to spoof a valid RFID tag and use it to send malicious data to the network.

\item	Sinkhole Attack: An adversary can compromise an IoT device (such as the root node) and perform sinkhole attack to claim exceptional capacity of computation, communication or power so that it could gain better chances for data forwarding by the neighboring nodes \cite{Adat}. Therefore, the attack allows an adversary to increase the amount of data before distributing it to the network. It is worthy to note that a sinkhole attack is the basic building block for an attacker to launch other harmful attacks in the network such as Denial of Service attack (DoS) \cite{rahman2016performance}. To overcome this attack, mechanism such as {\em secure routing protocols} are necessary to be studied and applied \cite{rahman2016simulation} \cite{Hollick}  \cite{fayaz2016effect}.

\item Man in the Middle Attack: As the name suggests, this attack allows an adversary controlling a malicious device to be {\em virtually positioned} between two transmitting nodes in IoT. Thus, the attacker can steal the identity information of the two surrounding devices, and acts as an intermediary to store and forward the communication that takes place between the two surrounding devices. It is worthy to note that the two devices cannot detect the presence of the attacker (i.e., they assume that a direct communication is taking place without the existence of a middleman). For example, consider a situation in which the adversary obtains the identifying information of the root and elected leader. This would seriously compromise the privacy, confidentiality and integrity of the LE process since sensitive LE messages can be monitored, eavesdropped and controlled by the adversary. To overcome this attack, {\em key management schemes}, {\em identity management techniques} and {\em secure routing protocols} are fundamental so that identity information are not leaked to malicious nodes in the network \cite{Das}.
\end{itemize}

We now provide an overview of the privacy issues, and the possible techniques to overcome them in IoT environment. In general, the data collected and used for LE purpose should go through the following three steps \cite{Zheng}:

\begin{itemize}
\item data collection
 \item aggregation, and
  \item data mining
 \end{itemize}

   In particular, data collection is used to sense and collect the related data of objects in IoT while data aggregation is intended to integrate only the {\em meaningful data} for LE (i.e., x-coordinates, identifiers, nodes’ residual energy and so on). However, the above process of LE in IoT does not ensure the privacy of objects. For instance, an adversary can obtain the confidential information such as identifier, location (i.e., coordinates) and the time when an election message is broadcasted in the network. Privacy, as a new challenge in IoT, can lead to serious issues such as the loss of property and endangering human safety. Thus, LE algorithms for IoT demands {\em privacy-preserving techniques} to ensure that the private data is not leaked to malicious parties in the network.

 Privacy-Preservation techniques are broadly classified into three types \cite{Hu}: 1) preserving privacy in the data collection phase : 2) preserving privacy in the data aggregation phase and 3) preserving privacy in the data mining phase. Mechanisms such as {\em key management} can be useful to preserve privacy at all level. However, most of the existing work on privacy in IoT is focused on the data aggregation phase.
Privacy preservation at the data aggregation stage is not a simple task because the relevant data needs to be processed in various {\em distinct locations} in the network. Thus, traditional encryption techniques are not suitable to achieve privacy at this stage. To solve this issue, various privacy-preserving techniques have been developed that exploit the data aggregation process. These techniques are broadly classified into three types \cite{Sen};
\begin{itemize}
    \item  Anonymity-based: Various anonymity-based techniques are proposed to preserve privacy at the data aggregation stage such as {\em T-closeness} and {\em K-anonymity}
    \item  Perturbation-based: These techniques are primarily concerned to perturb the raw data. To achieve this objective, various techniques are already proposed such as {\em random noise injection}, and {\em customization of data}.
    \item  Encryption-based: Different encryption techniques such as {\em zero-knowledge proof}, {\em homomorphic encryption}, and {\em secret sharing } can be used to preserve privacy in the data aggregation process
\end{itemize}

\section{Conclusion}
In this paper, we provided a systematic overview of the leader election algorithms in IoT, focusing on their essential features such as fault-tolerance, mobility, quality of service, data aggregation, synchronization together with security and privacy analysis. In addition, we also highlighted interesting open issues requiring further attention. We also proposed several new directions in terms of their essential features. We also showed the limitations of the current leader election algorithms in IoT, and possible techniques to overcome them. We believe that the work presented in this survey provides a solid foundation and new insights that would enable researchers to design more lightweight, fault tolerant, energy efficient and mobility-aware leader election algorithms in IoT.

%
%
\bibliographystyle{splncs04}
\bibliography{mybibliography}

\begin{thebibliography}{10}
\providecommand{\url}[1]{\texttt{#1}}
\providecommand{\urlprefix}{URL }
\providecommand{\doi}[1]{https://doi.org/#1}

\bibitem{Adat}
Adat, V., Gupta, B.: Security in internet of things: issues, challenges,
  taxonomy, and architecture. Telecommunication Systems  \textbf{67}(3),
  423--441 (2018)

\bibitem{Altisen}
Altisen, K., Cournier, A., Devismes, S., Durand, A., Petit, F.:
  Self-stabilizing leader election in polynomial steps. Information and
  Computation  \textbf{254},  330--366 (2017)

\bibitem{Awerbuch}
Awerbuch, B.: Optimal distributed algorithms for minimum weight spanning tree,
  counting, leader election, and related problems. In: Proceedings of the
  nineteenth annual ACM symposium on Theory of computing. pp. 230--240. ACM
  (1987)

\bibitem{Bello}
Bello, O., Zeadally, S.: Intelligent device-to-device communication in the
  internet of things. IEEE Systems Journal  \textbf{10}(3),  1172--1182 (2014)

\bibitem{Blin}
Blin, L., Tixeuil, S.: Compact self-stabilizing leader election for general
  networks. In: Latin American Symposium on Theoretical Informatics. pp.
  161--173. Springer (2018)

\bibitem{Bounceur2}
Bounceur, A., Bezoui, M., Euler, R., Kadjouh, N., Lalem, F.: Brogo: A new low
  energy consumption algorithm for leader election in wsns. In: 2017 10th
  International Conference on Developments in eSystems Engineering (DeSE). pp.
  218--223. IEEE (2017)

\bibitem{wbs}
Bounceur, A., Bezoui, M., Euler, R., Lalem, F.: A wait-before-starting
  algorithm for fast, fault-tolerant and low energy leader election in wsns
  dedicated to smart-cities and iot. In: 2017 IEEE SENSORS. pp.~1--3. IEEE
  (2017)

\bibitem{Bounceur3}
Bounceur, A., Bezoui, M., Euler, R., Lalem, F., Lounis, M.: A revised brogo
  algorithm for leader election in wireless sensor and iot networks. In: 2017
  IEEE SENSORS. pp.~1--3. IEEE (2017)

\bibitem{Bounceur1}
Bounceur, A., Bezoui, M., Lagadec, L., Euler, R., Abdelkader, L., Hammoudeh,
  M.: Dotro: A new dominating tree routing algorithm for efficient and
  fault-tolerant leader election in wsns and iot networks. In: International
  Conference on Mobile, Secure, and Programmable Networking. pp. 42--53.
  Springer (2018)

\bibitem{Chen}
Chen, W., Toueg, S., Aguilera, M.K.: On the quality of service of failure
  detectors. IEEE Transactions on computers  \textbf{51}(5),  561--580 (2002)

\bibitem{Chow}
Chow, Y.C., Luo, K.C., Newman-Wolfe, R.: An optimal distributed algorithm for
  failure-driven leader election in bounded-degree networks. In: Proceedings of
  the Third Workshop on Future Trends of Distributed Computing Systems. pp.
  136--141. IEEE (1992)

\bibitem{MCA}
Dagdeviren, O., Erciyes, K., Cokuslu, D.: Merging clustering algorithms in
  mobile ad hoc networks. In: International Conference on Distributed Computing
  and Internet Technology. pp. 56--61. Springer (2005)

\bibitem{Das}
Das, R., Singh, M., Majumder, K.: Sgsqot: A community-based trust management
  scheme in internet of things. In: Proceedings of International Ethical
  Hacking Conference 2018. pp. 209--222. Springer (2019)

\bibitem{Datta}
Datta, A.K., Devismes, S., Larmore, L.L., Villain, V.: Self-stabilizing weak
  leader election in anonymous trees using constant memory per edge. Parallel
  Processing Letters  \textbf{27}(02),  1750002 (2017)

\bibitem{Dolev}
Dolev, S., Israeli, A., Moran, S.: Uniform dynamic self-stabilizing leader
  election. IEEE Transactions on Parallel and Distributed Systems
  \textbf{8}(4),  424--440 (1997)

\bibitem{Effat}
EffatParvar, M., Yazdani, N., EffatParvar, M., Dadlani, A., Khonsari, A.:
  Improved algorithms for leader election in distributed systems. In: 2010 2nd
  International Conference on Computer Engineering and Technology. vol.~2, pp.
  V2--6. IEEE (2010)

\bibitem{fayaz2016effect}
Fayaz, M., Rahman, Z.U., Rahman, M.U., Abbas, S.: Effect of terrain size and
  pause time on the performance of reactive routing protocols. Jurnal Teknologi
   \textbf{78}(4-3) (2016)

\bibitem{IoT}
Gershenfeld, N., Krikorian, R., Cohen, D.: The internet of things. Scientific
  American  \textbf{291}(4),  76--81 (2004)

\bibitem{Hirschberg}
Hirschberg, D.S., Sinclair, J.B.: Decentralized extrema-finding in circular
  configurations of processors. Communications of the ACM  \textbf{23}(11),
  627--628 (1980)

\bibitem{Hollick}
Hollick, M., Nita-Rotaru, C., Papadimitratos, P., Perrig, A., Schmid, S.:
  Toward a taxonomy and attacker model for secure routing protocols. ACM
  SIGCOMM Computer Communication Review  \textbf{47}(1),  43--48 (2017)

\bibitem{Hu}
Hu, C., Luo, J., Pu, Y., Yu, J., Zhao, R., Huang, H., Xiang, T.: An efficient
  privacy-preserving data aggregation scheme for iot. In: International
  Conference on Wireless Algorithms, Systems, and Applications. pp. 164--176.
  Springer (2018)

\bibitem{Jan}
Jan, M.A., Nanda, P., He, X., Liu, R.P.: A sybil attack detection scheme for a
  forest wildfire monitoring application. Future Generation Computer Systems
  \textbf{80},  613--626 (2018)

\bibitem{Kim}
Kim, T.W., Kim, E.H., Kim, J.K., Kim, T.Y.: A leader election algorithm in a
  distributed computing system. In: Proceedings of the Fifth IEEE Computer
  Society Workshop on Future Trends of Distributed Computing Systems. pp.
  481--485. IEEE (1995)

\bibitem{Kuhn}
Kuhn, F., Oshman, R.: Dynamic networks: models and algorithms. ACM SIGACT News
  \textbf{42}(1),  82--96 (2011)

\bibitem{Kumar}
Kumar, A., Thomas, A.: Energy efficiency and network lifetime maximization in
  wireless sensor networks using improved ant colony optimization. Procedia
  engineering  \textbf{38},  3797--3805 (2012)

\bibitem{Larrea}
Larrea, M., Raynal, M.: Specifying and implementing an eventual leader service
  for dynamic systems. In: 2011 14th International Conference on Network-Based
  Information Systems. pp. 243--249. IEEE (2011)

\bibitem{Li}
Li, X., Niu, J., Kumari, S., Wu, F., Sangaiah, A.K., Choo, K.K.R.: A
  three-factor anonymous authentication scheme for wireless sensor networks in
  internet of things environments. Journal of Network and Computer Applications
   \textbf{103},  194--204 (2018)

\bibitem{spoofing}
Liu, D., Xu, Y., Huang, X.: Identification of location spoofing in wireless
  sensor networks in non-line-of-sight conditions. IEEE Transactions on
  Industrial Informatics  \textbf{14}(6),  2375--2384 (2017)

\bibitem{Lopez}
Lopez, J., Rios, R., Bao, F., Wang, G.: Evolving privacy: From sensors to the
  internet of things. Future Generation Computer Systems  \textbf{75},  46--57
  (2017)

\bibitem{Lu}
Lu, Y., Da~Xu, L.: Internet of things (iot) cybersecurity research: a review of
  current research topics. IEEE Internet of Things Journal  \textbf{6}(2),
  2103--2115 (2018)

\bibitem{Mondal}
Mondal, S., Ghosh, S., Dutta, P.: Energy efficient data gathering in wireless
  sensor networks using rough fuzzy c-means and aco. In: Industry Interactive
  Innovations in Science, Engineering and Technology, pp. 163--172. Springer
  (2018)

\bibitem{Ni}
Ni, Q., Pan, Q., Du, H., Cao, C., Zhai, Y.: A novel cluster head selection
  algorithm based on fuzzy clustering and particle swarm optimization. IEEE/ACM
  transactions on computational biology and bioinformatics  \textbf{14}(1),
  76--84 (2015)

\bibitem{Park}
Park, V.D., Corson, M.S.: A highly adaptive distributed routing algorithm for
  mobile wireless networks. In: Proceedings of INFOCOM'97. vol.~3, pp.
  1405--1413. IEEE (1997)

\bibitem{Pourghebleh}
Pourghebleh, B., Navimipour, N.J.: Data aggregation mechanisms in the internet
  of things: A systematic review of the literature and recommendations for
  future research. Journal of Network and Computer Applications  \textbf{97},
  23--34 (2017)

\bibitem{Rafailescu}
Rafailescu, M.: Fault tolerant leader election in distributed systems. arXiv
  preprint arXiv:1703.02247  (2017)

\bibitem{rahman2016simulation}
Rahman, M.U., Abbas, S.: Simulation-based analysis of manet routing protocols
  using group mobility model. In: 2016 International Conference on Inventive
  Computation Technologies (ICICT). vol.~1, pp.~1--5. IEEE (2016)

\bibitem{rahman2017lightweight}
Rahman, M.U., Abbas, S., Latif, S.: Lightweight detection of malicious nodes in
  mobile ad hoc networks. In: 2017 International Conference on Communication
  Technologies (ComTech). pp. 191--194. IEEE (2017)

\bibitem{rahman2016investigating}
Rahman, M.U., Alam, A., Abbas, S.: Investigating the impacts of entity and
  group mobility models in manets. In: 2016 International Conference on
  Computing, Electronic and Electrical Engineering (ICE Cube). pp. 181--185.
  IEEE (2016)

\bibitem{rahman2016performance}
Rahman, M.U., Rahman, Z.U., Fayaz, M., Abbas, S., ShahSani, R.K.: Performance
  analysis of tcp/aqm under low-rate denial-of-service attacks. In: 2016
  International Conference on Inventive Computation Technologies (ICICT).
  vol.~3, pp.~1--5. IEEE (2016)

\bibitem{Rawat}
Rawat, P., Singh, K.D., Chaouchi, H., Bonnin, J.M.: Wireless sensor networks: a
  survey on recent developments and potential synergies. The Journal of
  supercomputing  \textbf{68}(1),  1--48 (2014)

\bibitem{Reis}
Reis, V.A., Vieira, G.: Quality of service of an asynchronous crash-recovery
  leader election algorithm. arXiv preprint arXiv:1704.06302  (2017)

\bibitem{Sen}
Sen, A.A.A., Eassa, F.A., Jambi, K., Yamin, M.: Preserving privacy in internet
  of things: a survey. International Journal of Information Technology
  \textbf{10}(2),  189--200 (2018)

\bibitem{sharma}
Sharma, B., Bhatia, R.S., Singh, A.K.: A logical structure based fault tolerant
  approach to handle leader election in mobile ad hoc networks. Journal of King
  Saud University-Computer and Information Sciences  \textbf{29}(3),  378--398
  (2017)

\bibitem{Soundarabai}
Soundarabai, P.B., Sahai, R., Thriveni, J., Venugopal, K., Patnaik, L.:
  Efficient bully election algorithm in distributed systems (2013)

\bibitem{Tewari}
Tewari, A., Gupta, B.: Cryptanalysis of a novel ultra-lightweight mutual
  authentication protocol for iot devices using rfid tags. The Journal of
  Supercomputing  \textbf{73}(3),  1085--1102 (2017)

\bibitem{Tseng}
Tseng, P., Bertsekas, D.P., Tsitsiklis, J.N.: Partially asynchronous, parallel
  algorithms for network flow and other problems. SIAM Journal on Control and
  Optimization  \textbf{28}(3),  678--710 (1990)

\bibitem{Vasudevan}
Vasudevan, S., Kurose, J., Towsley, D.: Design and analysis of a leader
  election algorithm for mobile ad hoc networks. In: Proceedings of the 12th
  IEEE International Conference on Network Protocols, 2004. ICNP 2004. pp.
  350--360. IEEE (2004)

\bibitem{Yang}
Yang, X., Ren, X., Lin, J., Yu, W.: On binary decomposition based
  privacy-preserving aggregation schemes in real-time monitoring systems. IEEE
  Transactions on Parallel and Distributed Systems  \textbf{27}(10),
  2967--2983 (2016)

\bibitem{dynamic}
Yu, K., Gao, M., Jiang, H., Li, G.: Multi-leader election in dynamic sensor
  networks. EURASIP Journal on Wireless Communications and Networking
  \textbf{2017}(1), ~187 (2017)

\bibitem{Yu}
Yu, K., Gao, M., Jiang, H., Li, G.: Multi-leader election in dynamic sensor
  networks. EURASIP Journal on Wireless Communications and Networking
  \textbf{2017}(1), ~187 (2017)

\bibitem{Zheng}
Zheng, X., Cai, Z., Li, J., Gao, H.: Location-privacy-aware review publication
  mechanism for local business service systems. In: IEEE INFOCOM 2017-IEEE
  Conference on Computer Communications. pp.~1--9. IEEE (2017)

\end{thebibliography}

\end{document}